\documentclass{article}
\usepackage{spconf,amsmath,graphicx}
\usepackage{amsmath,graphicx}
\usepackage[center]{subfigure}
\usepackage{mathrsfs}
\usepackage{float}
\usepackage{multirow}
\usepackage{graphicx}
\usepackage{xcolor}
\usepackage{array}
\usepackage{mdwmath}
\usepackage{mdwtab}
\usepackage{eqparbox}
\usepackage{float, xcolor}
\usepackage{booktabs}
\usepackage{graphicx}
\usepackage{threeparttable}
\usepackage{url}
\usepackage{caption}

\title{Mixed-EVC: Mixed Emotion Synthesis and Control in Voice Conversion}
\name{Kun Zhou\textsuperscript{1}, Berrak Sisman\textsuperscript{2}, Carlos Busso\textsuperscript{2}, Bin Ma\textsuperscript{1}, Haizhou Li\textsuperscript{3,4}}
\address{
\textsuperscript{1}Speech Lab of DAMO Academy, Alibaba Group, Singapore\\
\textsuperscript{2}The University of Texas at Dallas, United States of America\\
\textsuperscript{3}The Chinese University of Hong Kong, Shenzhen, China
\textsuperscript{4}National University of Singapore, Singapore\\
}
\begin{document}
\ninept
\maketitle
\begin{abstract}
Emotional voice conversion (EVC) traditionally targets the transformation of spoken utterances from one emotional state to another, with previous research mainly focusing on discrete emotion categories. This paper departs from the norm by introducing a novel perspective: a nuanced rendering of mixed emotions and enhancing control over emotional expression.
To achieve this, we propose a novel EVC framework, Mixed-EVC, which only leverages discrete emotion training labels. We construct an attribute vector that encodes the relationships among these discrete emotions, which is predicted using a ranking-based support vector machine and then integrated into a sequence-to-sequence (seq2seq) EVC framework. Mixed-EVC not only learns to characterize the input emotional style but also quantifies its relevance to other emotions during training. As a result, users have the ability to assign these attributes to achieve their desired rendering of mixed emotions. Objective and subjective evaluations confirm the effectiveness of our approach in terms of mixed emotion synthesis and control while surpassing traditional baselines in the conversion of discrete emotions from one to another.
\end{abstract}
\section{Introduction}
\label{sec:intro}

%Consider bittersweet feelings that a human could express, 
Human speech often encompasses a blend of emotions, resulting in the emergence of complex emotional expressions, as evidenced in prior studies \cite{herzberg2012blend,williams2002can,chou2022exploiting}. Emotional voice conversion (EVC) aims to manipulate the emotional state of a spoken utterance while keeping speaker identity and linguistic content unchanged \cite{zhou2021emotional}. This paper represents a progressive step in the field of EVC, with a unique focus on infusing a quantifiable mixed emotion rendering into a human voice. The primary objective is to enhance the naturalness of human-computer interactions \cite{schuller2018age}, for example, enriching the emotional responses within a dialogue system \cite{amin2023wide,zhao2023chatgpt}.

%Voice conversion (VC) is the task of changing the speaker's identity while preserving the linguistic information \cite{sisman2020overview}. %Speaker identity primarily resides in the human vocal tract and manifests itself in spectral characteristics \cite{ramakrishnan2012speech}, making spectrum conversion the predominant focus of VC. 
%Compared to VC, EVC presents unique challenges \cite{kun2022emotion}. 

EVC poses unique challenges due to the complex structure of emotions \cite{russell2003core}. People use different words to describe the emotions that they feel, showing that there are nearly 34, 000 distinct emotions that a human may experience \cite{plutchik2001nature}. To understand how these emotions correlate with each other, scientists analyze them in a valence-arousal space \cite{russell2003core}. The evidence for the valence-arousal view comes from statistical analysis of how people report feelings \cite{ledoux2016we}. However, the analysis from a valence-arousal view has not always been able to tell us the real difference between emotions \cite{cowen2018many}. Plutchik's emotion wheel \cite{plutchik2013theories} provides a more straightforward way to describe emotions. 8 primary emotions: anger, fear, sadness, disgust, surprise, anticipation, trust, and joy, are arranged in an emotion wheel. The diverse amount of emotions could be produced either by changing the intensity or by adding up the primary emotions. Although preliminary studies \cite{tang23_interspeech, zhou2022speech} have explored synthesizing mixed emotions for text-to-speech systems, we observe a lack of study on mixed emotion synthesis in the literature of EVC, with existing studies mostly focusing on the conversion between discrete emotions. In this research, we draw inspiration from the emotion wheel theory and introduce an approach that employs voice conversion techniques to manipulate human emotions into a mixed emotional state.

Speech emotions are inherently supra-segmental and intricate, complex with multiple acoustic cues such as speech quality, pitch, energy, and speaking rate \cite{kun2022thesis}. Addressing these complexities in EVC calls for the modeling of both spectral and prosodic variations at the same time, leading to the study of sequence-to-sequence (seq2seq) architecture for EVC \cite{zhou21b_interspeech,robinson2019sequence,choi2021sequence,yang2022overview}.
%Speech emotion is inherently supra-segmental and complex with multiple acoustic cues, including speech quality, pitch, energy, and speaking rate \cite{xu2011speech}, where both spectral and prosodic variants need to be modeled. 
%Besides, the definition of emotions is subjective and ambiguous, which makes it difficult to precisely characterize emotions  \cite{cummins2019ambiguous}. 
%Voice conversion (VC) is the task of changing the speaker identity while preserving the linguistic information \cite{sisman2020overview}. Since speaker identity is characterized by the vocal tract and mainly manifested in spectrum \cite{ramakrishnan2012speech}, spectrum conversion has been the major focus of VC. 
%Compared to VC, EVC presents unique challenges. Speech emotion is inherently supra-segmental and complex with multiple acoustic cues, including speech quality, pitch, energy, and speaking rate \cite{xu2011speech}. Both spectral and prosodic variants need to be modeled in EVC. Besides, the definition of emotions is subjective and ambiguous, which makes it difficult to precisely characterize emotions  \cite{cummins2019ambiguous}. 
%Another challenge of EVC is the subjective and ambiguous nature of emotions makes emotions difficult to be precisely characterized \cite{cummins2019ambiguous}. 
To learn emotion information, existing EVC frameworks mostly leverage pre-defined discrete emotion labels as the supervision to emotion training, either
%There are generally two types of methods for EVC to model emotions in speech. 
learning a translation model between emotion pairs ~\cite{yamagishi2003modeling,aihara2012gmm,an2017emotional,Zhou2020,luo2019emotional}, or disentangling emotional elements with auto-encoders \cite{zhou2021vaw,zhou2020converting,kim2020emotional}. 
%Emotion conversion can be achieved by only manipulating emotional elements while keeping other speech characteristics unchanged. 
Emotion conversion can be achieved by assigning an emotion label or transferring from an emotional speech.
These methods restrict learning richer style descriptions of emotions but rather produce a stereotypical emotional pattern \cite{triantafyllopoulos2023overview}. Consequently, they confine emotions within specific categories, posing challenges when it comes to examining the connections between different emotional states, encompassing the entirety of human emotions, and creating a mixed emotional profile.
%which may not cover all human emotions, or create a mixed emotion profile. 
This paper aims to fill these gaps.
This paper presents the first investigation of mixed emotion synthesis and control for EVC, denoted as ``Mixed-EVC", which aims to address two challenges: (1) how to describe and quantify the combination of emotions; and (2) how to assess the produced mixed outcomes. Mixed-EVC introduces a novel approach to explicitly quantify and encode the relationships among discrete emotions into an attribute vector and distinguishes itself by infusing diverse emotional behaviors into the human voice leveraging limited discrete emotion labels. Mixed-EVC also allows users to intuitively and quantifiably control the emotion rendering through categorical classes, offering a more user-friendly alternative than manipulating continuous emotional attributes. Our key contributions can be outlined as follows:
\setlength{\leftmargini}{10pt} 
{\begin{itemize}
\setlength{\itemindent}{0pt}   %5. 最初のインデント
%\setlength{\labelsep}%{3pt}     %6. item と文字の間
    %\item This study is the first to model and synthesize mixed emotions for emotional voice conversion;
    \item We propose a novel formulation to quantify the mixture of emotions. We construct a ranking function for each pair of emotions, where the ranking values represent the degree of relevance with respect to an emotion; 
    %We regard the relevance as the attribute of the emotional style, and propose to explicitly model those attributes;
    \item During training, the attribute can be precisely predicted by the ranking function, guiding the decoder to quantify the relevance between the input emotional style and all other emotions. At run-time, the users can define those attributes to generate various emotional mixtures;
    \item We design evaluation metrics to assess the effectiveness of our approach in terms of synthesizing mixed emotions and enabling precise control.
    %present an interactive demo\footnote{\textbf{Emotion Triangle:} \url{https://demo9646.github.io/Emotion_Triangle/}} to the readers.
\end{itemize}}
\vspace{-2mm}

\begin{figure}
    \centering
    \includegraphics[width=0.5\textwidth]{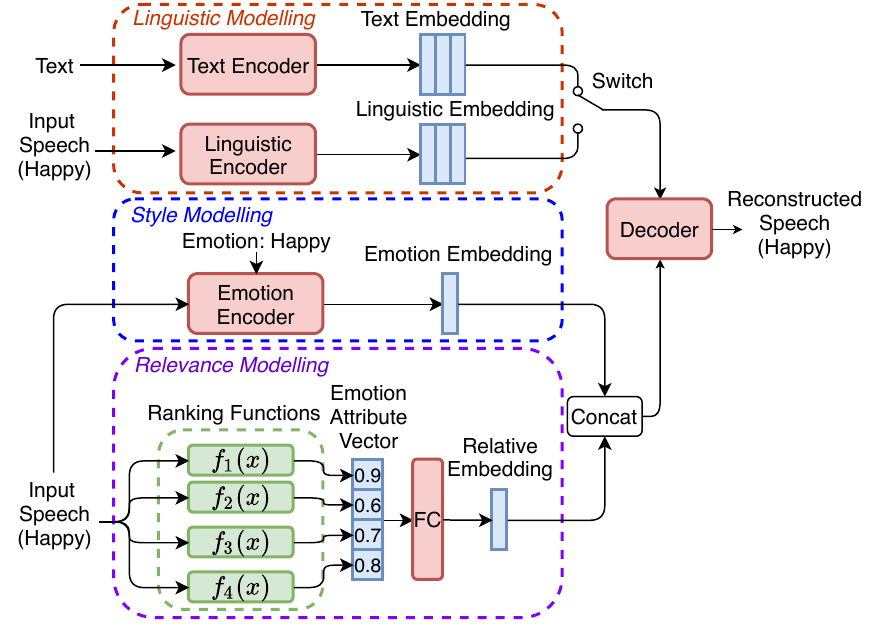}
    \vspace{-3mm}
    \caption{The training diagram of Mixed-EVC, where the red boxes represent the modules that are involved in the training. The pre-trained ranking functions automatically predict emotion attributes that indicate the degree of relevance between the input emotional style (Happy) and other emotional styles (Angry, Sad, Surprise and Neutral).}
    \label{fig:train}
    \vspace{-5mm}
\end{figure}

\section{Mixed-EVC}
%We propose a sequence-to-sequence emotional voice conversion framework that allows for 
In this section, we present the details of our proposed emotional voice conversion framework, Mixed-EVC. We begin by introducing a novel approach that facilitates the synthesis of mixed emotions (Sec. 2.1). Following that, we introduce the core components and the training diagram of Mixed-EVC (Sec. 2.2). Lastly, we describe how to render and control the desired mixed emotions (Sec. 2.3).
\vspace{-2mm}
\subsection{Modeling Emotion Attribute with Pairwise Ranking}
Most emotional speech databases group utterances into several discrete categories thus restricting to learning of richer descriptions of the emotional style \cite{zhou2021emotional}. 
Instead of predicting the presence of a specific emotion, we propose to model
the relative difference between a pair of emotions, which we define as \textit{``Emotion Attribute"}.  During the training, we characterize the input emotional style by measuring its relevance to other emotions by explicitly modeling emotion attributes. For example, while it is hard to give a consensus emotion label to an utterance, especially when the emotions are ambiguous, we can agree that it sounds less happy than A but angrier than B \cite{yannakakis2018ordinal}. 

We employ the idea of relative attributes \cite{parikh2011relative} to encode the relationship between discrete emotions into a quantifiable vector. Relative attributes were first proposed in computer vision \cite{parikh2011relative} and later applied in speech processing \cite{zhu2019controlling,lei2021fine,zhou2022emotion}. Prior studies show that relative attributes provide a more detailed description of an image, thus surpassing traditional classifiers on image recognition tasks \cite{parikh2011relative}.  
%Similar to relative attributes \cite{parikh2011relative} that are studied in computer vision and later applied in speech processing \cite{zhu2019controlling,lei2021fine,zhou2022emotion}, 
Inspired by that, we propose to learn a ranking function for each emotion attribute given relative similarity constraints on paired emotional samples. More specifically, we are given a training set $T = \{\mathbf{x}_n\}$, where ${\mathbf{x}_n}$ is the acoustic feature vector of the $n^{th}$ training sample with $M$ emotion labels.
Our goal is to learn a ranking function $f_m$ given below:
\begin{equation}
    f_m(\mathbf{x}_n) = \mathbf{W}_m\mathbf{x}_n, 
\end{equation}
where $m = 1, ..., M$ and $\mathbf{W}$ is a weighting matrix.
We construct an ordered set $O_m$ and an unordered set $U_m$ as the supervision sets, which satisfy the following constraints:
\begin{align}
    \forall (x_i, x_j) \in O_m: \mathbf{W}_m\mathbf{x}_i > \mathbf{W}_m\mathbf{x}_j\\
    \forall (x_i, x_j) \in U_m: \mathbf{W}_m\mathbf{x}_i = \mathbf{W}_m\mathbf{x}_j, 
\end{align}
The sample $x_i$ should have a stronger presence of emotion attribute $a_m$ than $x_j$ in the ordered set $O_m$, while their presence should be similar in the unordered set $U_m$.
The weighting matrix $\mathbf{W}$ is estimated by solving a support vector machine (SVM) problem \cite{chapelle2007training}.
%\begin{align}
%    \min_{\mathbf{W}_m} (\frac{1}{2} \parallel \mathbf{W}_m \parallel_2^2 + C(\sum \xi_{i,j}^2 + \sum \gamma_{i,j}^2))\\
%    \text{ s.t. }           \mathbf{W}_m(\mathbf{x}_i - \mathbf{x}_j) \geq 1 - \xi_{i,j}; \forall(i,j) \in O_m\\
%    |\mathbf{W}_m(\mathbf{x}_i-\mathbf{x}_j)|\leq \gamma_{i,j}; \forall(i,j) \in U_m\\
%    \xi_{i,j} \geq 0; \gamma_{i,j}\geq 0, 
%\end{align}
%where $C$ is the trade-off between the margin and the size of slack variables $\xi_{i,j}$ and $\gamma_{i,j}$. The emotion attribute is the distance to the decision boundary. We create pairs for all the emotions and repeat the aboveranking training in a pairwise manner. At run-time, the trained ranking function can automatically predict the emotion attribute that indicates the relevance or similarity between a pair of emotional styles. In practice, each emotion attribute is normalized into [0,1], where a smaller value represents a more similar emotion style. All the emotion attributes form an emotion attribute vector that measures the degree of relevance of an input emotional style to all other emotions. 
%{It is noted that normalization does not have any effects because the values represent the order rather than the actual distance.}
%where $C$ represents the trade-off factor between the margin and the scale of slack variables $\xi_{i,j}$ and $\gamma_{i,j}$. 
The emotion attribute thus corresponds to the distance from the decision boundary. We generate pairs for all emotions and iteratively conduct the aforementioned ranking training in a pairwise manner. During inference, the trained ranking function can automatically predict the emotion attribute, which signifies the relevance or similarity between two emotional styles. Each emotion attribute is standardized within the [0, 1] range to be easily quantifiable, where a lower value indicates a greater similarity in emotional style. All these emotion attributes together compose an emotion attribute vector, quantifying the degree of relevance between an input emotional style and all other emotions. Next, we explain how those attributes contribute to the EVC training. 
\begin{figure}
    \centering
    \includegraphics[width=0.5\textwidth]{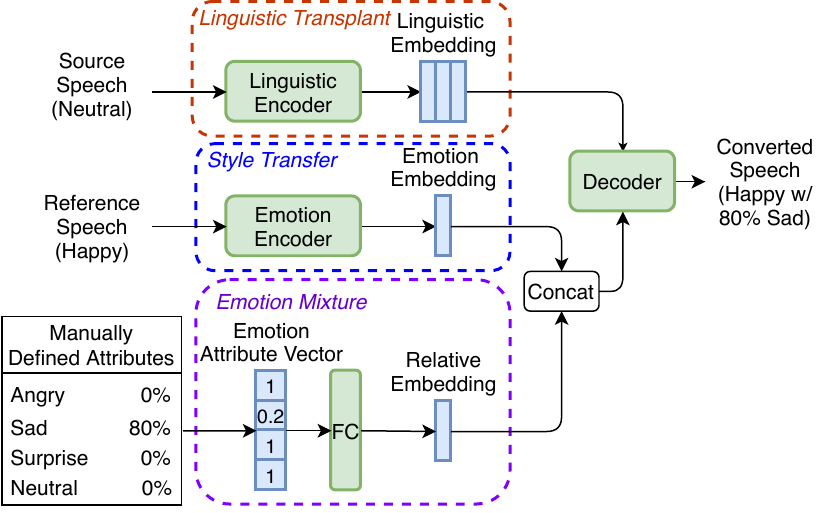}
    \vspace{-3mm}
    \caption{The run-time diagram of the proposed Mixed-EVC, where the green boxes represent pre-trained modules. By assigning manually defined attributes, Mixed-EVC transfers the source emotion to a target emotional mixture while preserving the source linguistic content.}
    \label{fig:run-time}
    \vspace{-4mm}
\end{figure}

\begin{figure*}[t]
\centering
\subfigure[Mixing Angry with Surprise \newline (Outrage)]
{
\begin{minipage}[c]{0.235\linewidth}
\centering
\includegraphics[width=1\textwidth]{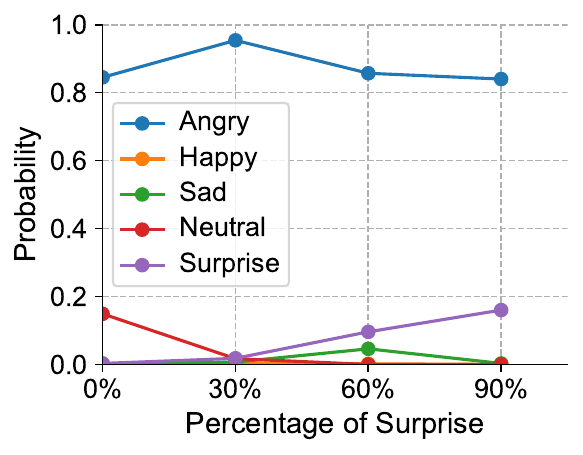}
\end{minipage}%
}
\subfigure[ Mixing Happy with Surprise \newline (Excitement)]
{
\begin{minipage}[c]{0.235\linewidth}
\centering
\includegraphics[width=1\textwidth]{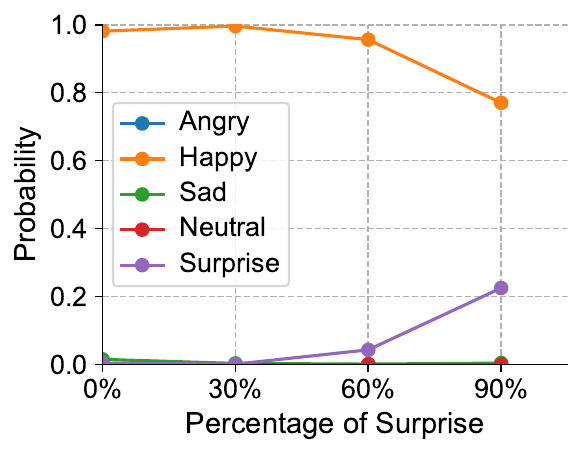}
\end{minipage}
}
\subfigure[Mixing Sad with Surprise \newline (Disappointment)]
{
\begin{minipage}[c]{0.235\linewidth}
\centering
\includegraphics[width=1\textwidth]{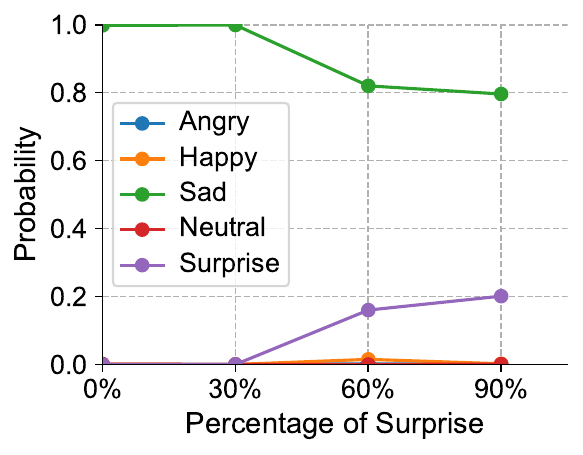}
\end{minipage}
}
\subfigure[Mixing Happy with Sad \newline (Bittersweet)]
{
\begin{minipage}[c]{0.235\linewidth}
\centering
\includegraphics[width=1\textwidth]{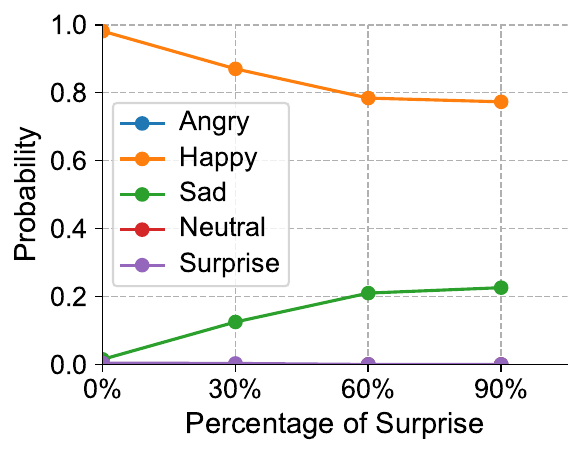}
\end{minipage}
}
\centering
\vspace{-4mm}
\caption{The classification probabilities utilizing a pre-trained SER model. Each data point corresponds to an averaged probability value obtained by analyzing 20 converted utterances displaying mixed emotions, all of which are drawn from the evaluation set.}
\vspace{-4mm}
\label{fig:ser}
\end{figure*}
\subsection{Seq2Seq Emotional Voice Conversion Training}

We integrate our emotion attribute vector into a seq2seq-based emotional voice conversion framework, as illustrated in Figure \ref{fig:train}. Our approach, distinct from traditional frame-wise modeling \cite{yang2022overview}, jointly models feature mapping and alignment and automatically predicts speech duration during run-time.

Given an input speech, the linguistic encoder is trained to predict a sequence of linguistic embeddings (``Linguistic Modeling"), while the emotion encoder focuses on encoding the emotional style into an utterance-level emotion embedding (``Style Modeling"). The pre-trained relative functions further predict an emotion attribute vector that signifies the relevance of the input emotional style to other emotions (``Relevance Modeling"). This emotion attribute vector is then transformed using a fully connected (FC) layer, resulting in a relative embedding. Ultimately, the decoder learns to reconstruct the input emotional style by combining information from both the emotion and relative embeddings. Our framework not only explicitly characterizes the input emotional style but also establishes a relationship with other emotions.

To overcome the instability issues in seq2seq models, we conduct emotional training with the following strategies: (1) introducing text supervision \cite{zhang2019non}, and (2) pre-training with a large neutral-speaking corpus \cite{zhou21b_interspeech}. We use text transcriptions to assist the framework in disentangling the linguistic information from the speech. A text encoder predicts a sequence of text embeddings from the input text. The text and linguistic embeddings are then fed into the decoder in an alternative manner. {Following the previous literature \cite{zhang2019non}, we employ a contrastive loss to ensure the similarity between text and linguistic embeddings. 
%We also employ adversarial training with an emotion classifier to get more disentangled linguistic representations. The style encoder learns speaker style with the supervision of one-hot speaker labels. 

\subsection{Mixed Emotion Rendering and Control}

%At run-time, Mixed-EVC converts the source emotion into an emotional mixture, as shown in Figure \ref{fig:run-time}. The linguistic encoder converts the source linguistic content into internal representations. The emotion encoder encapsulates a set of reference speech into an emotion embedding. The characteristics of other emotion types can be further introduced by manually defining an attribute vector. It allows us to vary the percentage for each emotion type and create different emotional mixtures. 

During run-time, Mixed-EVC carries out the conversion of the source emotion into an emotional mixture, illustrated in Figure \ref{fig:run-time}. This process involves several steps: first, the linguistic encoder transforms the source linguistic content into internal representations. Next, the emotion encoder encapsulates a set of reference speech into an utterance-level emotion embedding. Additionally, attributes of other emotion types can be introduced by manually specifying an attribute vector. This flexibility enables us to adjust the proportions of each emotion type, thereby enabling the creation of diverse emotional mixtures.

\vspace{-3mm}
\section{Experiments}

\subsection{Experimental Setup}
%We report two case studies including objective and subjective evaluations to validate our idea of mixed emotion synthesis.
We conduct experiments with the ESD dataset \cite{zhou2021emotional}, where we randomly choose one female speaker (``0019") with five emotions (\textit{``Neutral"}, \textit{``Happy"}, \textit{``Sad"}, \textit{``Angry"} and \textit{``Surprise"}). We follow the data partition in the ESD dataset, and for each emotion, we use 300, 30, and 20 utterances for training, testing, and evaluation, respectively. We train a universal EVC model for all emotions. At run-time, we convert \textit{Neutral} to 4 different mixtures of emotions that are: 
\begin{itemize}
    \item \textbf{Outrage}, \textbf{Excitement}, \textbf{Disappointment}: where we convert \textit{Neutral} to \textit{Angry}, \textit{Happy} and \textit{Sad} (Neu-Ang, Neu-Hapy, Neu-Sad) respectively while introducing different percentages of \textit{Surprise} into the mixture. These 3 mixtures have been studied in emotion theory \cite{plutchik2013theories};
    \item \textbf{Bittersweet}: where we convert \textit{Neutral} to \textit{Happy} while introducing different percentages of \textit{Sad}. Even though happiness and sadness are two oppositely valenced emotions in Russell's model \cite{russell1980circumplex}, there are some debates that agree with the co-existence of happiness and sadness \cite{williams2002can};
\end{itemize}

The training pipeline is described below: We first pre-train the relative ranking function between each emotion pair. Each utterance is represented by a 384-dimensional feature vector defined by the INTERSPEECH Emotion Challenge \cite{schuller2009interspeech}, which is used to train the ranking functions. The ranking functions achieved 98\% accuracy in classifying the emotion categories on the test set of the ESD dataset.
We then follow a 2-stage training strategy \cite{zhou21b_interspeech} to train our proposed framework: (1) Style Pre-training with the VCTK Corpus \cite{veaux2016vctk}, and (2) Seq2Seq EVC Training with the ESD dataset. The inputs to the EVC model are acoustic features represented by an 80-dimensional Mel-spectrogram extracted every $12.5$\, ms with a frame size of $50$\, ms for the short-time Fourier transform (STFT) and phoneme sequences converted by the Festival \cite{black2001festival} G2P tool. It should be noted that we only use acoustic features as inputs during the conversion.

Our proposed framework has a similar structure to the model presented in  \cite{zhang2019non}. The linguistic encoder comprises an encoder, a 2-layer 256-cell BLSTM, and a decoder employing a 1-layer 512-cell BLSTM with an attention layer followed by a fully connected (FC) layer with an output channel of $512$. The decoder has the same model architecture as that of Tacotron \cite{wang2017tacotron}. The text encoder is a 3-layer 1D CNN with a kernel size of $5$ and a channel of $512$. 
The style encoder is a 2-layer, 128-cell BLSTM followed by an FC layer with an output channel of $64$. The classifier is a 4-layer FC with $512$, $512$, $512$, and $5$ channels. 
%We set the batch size to $64$ and $4$ for style pre-training and emotion training, respectively. We set the learning rate for style pre-training at $0.001$ and the weight decay at $0.0001$; for emotion pre-training, we halve the learning rate every 7 epochs.
For training, we specify a batch size of 64 for style pre-training and 4 for emotion training. During style pre-training, the learning rate is set at 0.001, and a weight decay of 0.0001 is applied. For emotion pre-training, we adopt a learning rate halving strategy every 7 epochs.
\subsection{Objective Evaluation}

%We evaluate our synthesized mixed emotions with a speech emotion recognition (SER) model that is pre-trained on the ESD dataset. The SER model has a similar structure to that of \cite{chen20183}, which consists of a 3-D CNN layer, a BLSTM layer,  an attention layer, and an FC layer. We use the classification probabilities derived from the last softmax layer of the SER to analyze the performance of mixed emotions. We believe that classification probabilities summarize the emotional information from previous layers for decision-making and provide us with a tool to study emotional mixtures.

We assess the quality of our synthesized mixed emotions by employing a speech emotion recognition (SER) model pre-trained on the ESD dataset. This SER model shares a similar structure to the one presented in \cite{chen20183}, encompassing components such as a 3-D CNN layer, a BLSTM layer, an attention layer, and an FC layer. Our analysis leverages the classification probabilities obtained from the final softmax layer of the SER. We contend that these classification probabilities serve as a concise summary of emotional cues aggregated from preceding layers, aiding in decision-making and providing us with a valuable tool for investigating emotional blends.

We report the classification probabilities in Figure \ref{fig:ser}. We first evaluate three different mixed effects that are \textit{Outrage}, \textit{Excitement} and \textit{Disappointment}, where we convert \textit{Neutral} to \textit{Angry}, \textit{Neutral} to \textit{Happy} and \textit{Neutral} to \textit{Sad} respectively. These transformations are done while gradually increasing the percentage (0\%, 30\%, 60\% and 90\%) of \textit{Surprise}. As shown in Figure \ref{fig:ser}(a), (b), and (c), we observe that the probability of \textit{Surprise} consistently increases when we increase the percentage of \textit{Surprise} from 0\% to 90\%. The likelihoods of \textit{Angry}, \textit{Happy}, and \textit{Sad} consistently maintain the highest in three different emotional mixtures. This pattern arises from their distinct characterization by the emotion encoder, establishing them as the predominant emotional components within the mixture. We then evaluate the mixed effect of \textit{Bittersweet} as shown in Figure \ref{fig:ser}(d), where \textit{Sad} is further introduced when we convert \textit{Neutral} to \textit{Happy}. As shown in Figure \ref{fig:ser}(d), we find a similar observation as in Figure \ref{fig:ser}(a), (b) and (c). These observations indicate that mixed emotions can be objectively recognized by a pre-trained SER.

\begin{table}[t]
\centering
\caption{MOS with 95\% confidence interval to evaluate the speech quality of synthesized mixed emotions.}
\vspace{+2mm}
\scalebox{0.8}{
\begin{tabular}{cc|c}
\toprule
\multicolumn{2}{c}{\textbf{Configuration}}                                                                                                         & \textbf{MOS}  \\ \midrule
\multicolumn{1}{c|}{\multirow{5}{*}{\begin{tabular}[c]{@{}c@{}}Mixing Angry \\with Surprise\end{tabular}}} & Ground truth (Angry) & 4.78 $\pm$ 0.17 \\ %\cmidrule(l){2-3} 
\multicolumn{1}{c|}{}                                                                                               & + 0\% Surprise          & 3.50 $\pm$ 0.24 \\ %\cmidrule(l){2-3} 
\multicolumn{1}{c|}{}                                                                                               & + 30\% Surprise         & 3.34 $\pm$ 0.26 \\ %\cmidrule(l){2-3} 
\multicolumn{1}{c|}{}                                                                                               & + 60\% Surprise         & 3.15 $\pm$ 0.34 \\ %\cmidrule(l){2-3} 
\multicolumn{1}{c|}{}                                                                                               & + 90\% Surprise        & 3.20 $\pm$ 0.31 \\ \midrule
\multicolumn{1}{c|}{\multirow{5}{*}{\begin{tabular}[c]{@{}c@{}}Mixing Happy\\ with Surprise\end{tabular}}} & Ground truth (Happy) & 4.85 $\pm$ 0.12 \\ %\cmidrule(l){2-3} 
\multicolumn{1}{c|}{}                                                                                               & + 0\% Surprise          & 3.63 $\pm$ 0.27 \\ %\cmidrule(l){2-3} 
\multicolumn{1}{c|}{}                                                                                               & + 30\% Surprise        & 3.53 $\pm$ 0.19 \\ %\cmidrule(l){2-3} 
\multicolumn{1}{c|}{}                                                                                               & + 60\% Surprise         & 3.17 $\pm$ 0.30 \\ %\cmidrule(l){2-3} 
\multicolumn{1}{c|}{}                                                                                               & + 90\% Surprise         & 3.05 $\pm$ 0.34 \\ \midrule
\multicolumn{1}{c|}{\multirow{5}{*}{\begin{tabular}[c]{@{}c@{}}Mixing Sad \\ with Surprise\end{tabular}}}   & Ground truth (Sad)   & 4.74 $\pm$ 0.12 \\ %\cmidrule(l){2-3} 
\multicolumn{1}{c|}{}                                                                                               & + 0\% Surprise           & 3.63 $\pm$ 0.27 \\ %\cmidrule(l){2-3} 
\multicolumn{1}{c|}{}                                                                                               & + 30\% Surprise           & 3.53 $\pm$ 0.19 \\ %\cmidrule(l){2-3} 
\multicolumn{1}{c|}{}                                                                                               & + 60\% Surprise           & 3.17 $\pm$ 0.30 \\ %\cmidrule(l){2-3} 
\multicolumn{1}{c|}{}                                                                                               & + 90\% Surprise           & 3.05 $\pm$ 0.34  \\\midrule
\multicolumn{1}{c|}{\multirow{5}{*}{\begin{tabular}[c]{@{}c@{}}Mixing Happy \\ with Sad\end{tabular}}}   & Ground truth (Happy)   & 4.84 $\pm$ 0.12 \\ %\cmidrule(l){2-3} 
\multicolumn{1}{c|}{}                                                                                               & + 0\% Sad            & 3.17 $\pm$ 0.35 \\ %\cmidrule(l){2-3} 
\multicolumn{1}{c|}{}                                                                                               & + 30\% Sad           & 3.48 $\pm$ 0.30 \\ %\cmidrule(l){2-3} 
\multicolumn{1}{c|}{}                                                                                               & + 60\% Sad           & 3.42 $\pm$ 0.32 \\ %\cmidrule(l){2-3} 
\multicolumn{1}{c|}{}                                                                                               & + 90\% Sad           & 3.08 $\pm$ 0.36  \\\bottomrule
\end{tabular}}
\label{tab: mos}
\vspace{-3mm}
\end{table}

\begin{table}[t]
%\vspace{-5mm}
\centering
\caption{BWS test results to evaluate the perception of the mixed feelings (\textit{Outrage}, \textit{Excitement}, \textit{Disappointment} and \textit{Bittersweet}) in the converted mixed emotions.}
\vspace{+2mm}
\scalebox{0.8}{
\begin{tabular}{cl|l|l}
\toprule
\multicolumn{2}{c|}{\textbf{Configuration}}                                                                                                        & \multicolumn{1}{c|}{\textbf{Best (\%)}} & \textbf{Worst (\%)} \\ \midrule
\multicolumn{4}{c}{\textbf{(a) Perception of Outrage}}                                                                                                                                                  \\ \midrule
\multicolumn{1}{c|}{\multirow{4}{*}{\begin{tabular}[c]{@{}c@{}}Mixing Angry \\ with Surprise\end{tabular}}} & \multicolumn{1}{c|}{+ 0\% Surprise} &  \multicolumn{1}{c|}{15.4}              &    \multicolumn{1}{c}{40.0}             \\ 
\multicolumn{1}{c|}{}                                                                                       & \multicolumn{1}{c|}{+ 30\% Surprise} & \multicolumn{1}{c|}{15.4}              &     \multicolumn{1}{c}{24.6}           \\ 
\multicolumn{1}{c|}{}                                                                                       & \multicolumn{1}{c|}{+ 60\% Surprise} & \multicolumn{1}{c|}{13.8}              &     \multicolumn{1}{c}{21.6}           \\ 
\multicolumn{1}{c|}{}                                                                                       & \multicolumn{1}{c|}{+ 90\% Surprise} & \multicolumn{1}{c|}{55.4}              &     \multicolumn{1}{c}{13.8}           \\ \midrule
\multicolumn{4}{c}{\textbf{(b) Perception of Excitement}}                                                                                                                                               \\ \midrule
\multicolumn{1}{c|}{\multirow{4}{*}{\begin{tabular}[c]{@{}c@{}}Mixing Happy\\ with Surprise\end{tabular}}}  & \multicolumn{1}{c|}{+ 0\% Surprise}  & \multicolumn{1}{c|}{3.1}              &     \multicolumn{1}{c}{69.2}           \\  
\multicolumn{1}{c|}{}                                                                                       & \multicolumn{1}{c|}{+ 30\% Surprise} & \multicolumn{1}{c|}{12.3}              &    \multicolumn{1}{c}{10.8}            \\ 
\multicolumn{1}{c|}{}                                                                                       & \multicolumn{1}{c|}{+ 60\% Surprise} & \multicolumn{1}{c|}{23.1}              &    \multicolumn{1}{c}{9.2}            \\ 
\multicolumn{1}{c|}{}                                                                                       & \multicolumn{1}{c|}{+ 90\% Surprise} & \multicolumn{1}{c|}{61.5}              &    \multicolumn{1}{c}{10.8}             \\ \midrule
\multicolumn{4}{c}{\textbf{(c) Perception of Disappointment}}                                                                                                                                           \\ \midrule
\multicolumn{1}{c|}{\multirow{4}{*}{\begin{tabular}[c]{@{}c@{}}Mixing Sad\\ with Surprise\end{tabular}}}    & \multicolumn{1}{c|}{+ 0\% Surprise}  & \multicolumn{1}{c|}{13.8}              &     \multicolumn{1}{c}{53.8}            \\ 
\multicolumn{1}{c|}{}                                                                                       & \multicolumn{1}{c|}{+ 30\% Surprise} & \multicolumn{1}{c|}{16.9}              &     \multicolumn{1}{c}{12.3}            \\ 
\multicolumn{1}{c|}{}                                                                                       & \multicolumn{1}{c|}{+ 60\% Surprise} & \multicolumn{1}{c|}{21.5}              &     \multicolumn{1}{c}{27.7}            \\  
\multicolumn{1}{c|}{}                                                                                       & \multicolumn{1}{c|}{+ 90\% Surprise}  & \multicolumn{1}{c|}{47.8}              &      \multicolumn{1}{c}{6.2}           \\ \midrule
\multicolumn{4}{c}{\textbf{(d) Perception of Bittersweet}}                                                                                                                                              \\ \midrule
\multicolumn{1}{c|}{\multirow{4}{*}{\begin{tabular}[c]{@{}c@{}}Mixing Happy\\ with Sad\end{tabular}}}       & \multicolumn{1}{c|}{+ 0\% Sad}       & \multicolumn{1}{c|}{4.6}              &      \multicolumn{1}{c}{41.5}           \\ 
\multicolumn{1}{c|}{}                                                                                       & \multicolumn{1}{c|}{+ 30\% Sad}      & \multicolumn{1}{c|}{20.0}              &      \multicolumn{1}{c}{12.3}           \\ 
\multicolumn{1}{c|}{}                                                                                       & \multicolumn{1}{c|}{+ 60\% Sad}      & \multicolumn{1}{c|}{30.8}              &    \multicolumn{1}{c}{10.8}             \\ 
\multicolumn{1}{c|}{}                                                                                       & \multicolumn{1}{c|}{+ 90\% Sad}      & \multicolumn{1}{c|}{44.6}              &     \multicolumn{1}{c}{35.4}            \\ \bottomrule
\end{tabular}}
\label{tab:xbs}
%\vspace{-5mm}
\end{table}

\subsection{Subjective Evaluation}

We then conduct listening experiments to evaluate our synthesized results in terms of speech quality and emotion perception. 15 subjects participated in all the experiments and each of them listened to 112 synthesized utterances in total. 

We first report mean opinion scores (MOS) results for speech quality, where all participants are asked to listen to the reference speech (``Ground truth'') and the synthesized speech with mixed emotions and score the ``quality'' of each speech sample on a 5-point scale (`5' for excellent, `4' for good, `3' for fair, `2' for poor, and `1' for bad). As shown in Table \ref{tab: mos}, the audio quality of synthesized speech slightly decreases as we increase the percentage of emotional mixture. Nonetheless, the quality remains between fair and good across the board.

We then conduct best-worst scaling (BWS) tests to evaluate the emotion perception of the synthesized mixed emotions. All participants are asked to choose the best and the worst emotion according to their perception of the mixed emotion (\textit{Outrage}, \textit{Excitement}, \textit{Disappointment} and \textit{Bittersweet}). As shown in Table \ref{tab:xbs}(a), (b) and (c), we observe that most participants can perceive the mixed feelings and choose those with 90\% of \textit{Surprise} as the ``Best", and those of 0\% of \textit{Surprise} as the ``Worst". We take one step further to evaluate the perception of \textit{Bittersweet} as shown in Table \ref{tab:xbs}(d). Most participants chose those with 90\% \textit{Sad} as the ``Best" confirming the strength of our approach. The results of \textit{Bittersweet} are not as distinguishable as those of the other three mixed emotions, which suggests that it could be more challenging for listeners to perceive a \textit{Bittersweet} feeling. These results show that we are able to create new emotional feelings that are subtle but do not exist in the database. We also show the effectiveness of controllability by varying the percentages of primary emotions to synthesize different emotional mixtures.

{We further conduct a preference test to evaluate the performance of Mixed-EVC on one-to-one emotion conversion. We choose Seq2Seq-EVC \cite{zhou21b_interspeech} as the baseline. Participants are asked to choose the best one in terms of the emotion similarity with the reference emotion. As shown in Figure \ref{fig:preference}, our proposed Mixed-EVC outperforms the baseline Seq2Seq-EVC in all conversion pairs. These results show that Mixed-EVC could not only synthesize emotional mixtures but also have superior performance on emotion conversion.}
\begin{figure}[t]
    \centering
    \includegraphics[width=0.38\textwidth, trim={0 0 0 13mm},clip]{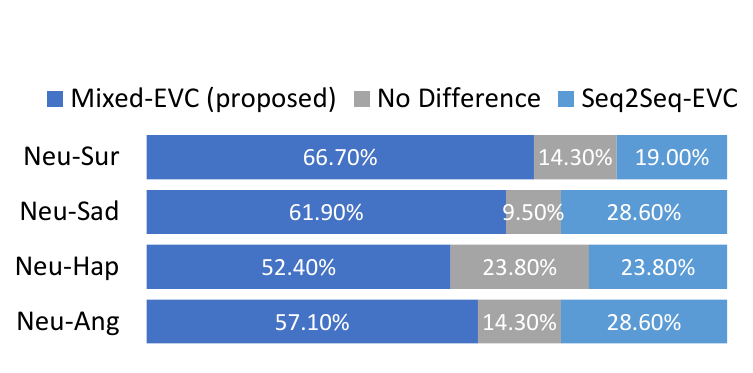}
    \vspace{-5mm}
    \caption{Prereference test results of emotion similarity between the proposed Mixed-EVC and the baseline Seq2Seq-EVC over one-to-one emotion conversion.}
    \label{fig:preference}
    \vspace{-5mm}
\end{figure}
%\subsection{An Interactive Demo: Emotion Triangle}
%Mixed emotions can be perceived during the emotion transition. As an application study, we built an emotion triangle for users to experience the emotion transitions, where the key challenge is to model the internal states between emotions. With our proposed Mixed-EVC, it is possible for us to synthesize those states by mixing with different percentages of emotions. 
%Readers are suggested to refer to the demo page. 
\vspace{-3mm}
\section{Conclusion}
We introduce Mixed-EVC, a seq2seq emotional voice conversion (EVC) framework, to address the existing research gap in mixed emotion synthesis within the context of EVC.  We formulate emotional styles as an attribute and explicitly model the degree of relevance between different emotions through a ranking-based SVM. By manually adjusting relevance at run-time, Mixed-EVC could produce different emotional mixtures. Both objective and subjective evaluations show the effectiveness of synthesizing different mixed emotions. The speech samples are publicly available \footnote{\textbf{Speech Demo}: \url{https://demo9646.github.io/Mixed_EVC/}}.
%We further present emotion transition within an emotion triangle.

\footnotesize

\bibliographystyle{IEEEtran}
\bibliography{mybib}

\end{document}